\documentclass[11pt,twoside]{article}
\usepackage{cozumel2005}
\usepackage{epsf}
\usepackage{psfig}
\usepackage{lscape}
\begin{document}
\title{Two New Milky Way Companions}    
\author{Beth Willman}   
\affil{NYU, Center for Cosmology and Particle Physics}    

\begin{abstract} 

We discuss the detection limits and current status of a uniform survey
of SDSS I for ultra-faint Milky Way dwarf galaxies.  We present the
properties of two new, low surface brightness Milky Way companions
discovered as a result of this survey.  One of these companions is the
Ursa Major dwarf, the newest dwarf spheroidal companion to the Milky
Way and the lowest luminosity galaxy yet known.  Ursa Major is
about 100 kpc away and is similar to Sextans, but with roughly an
order of magnitude fewer stars.  The other companion, SDSSJ1049+5103,
lies $\sim$ 50 kpc away.  Its stellar distribution suggests that it
may be undergoing tidal stripping. This companion is extremely faint
(M$_V$ $\sim$ -3) but has a large half-light size for its luminosity.  It is
therefore unclear whether it is a globular cluster or a dwarf galaxy.
\end{abstract}

\section{Introduction}                      
Of all galaxies that have survived until the present epoch, the lowest
mass dwarf galaxies inhabit the dark matter halos with the shallowest
potential wells and have been the most limited in their ability to
cool gas.  As such, their properties are the most sensitive to
physical processes that control galaxy formation.  Most of the least
massive dwarfs currently cataloged have been found near the Milky Way.
Milky Way dwarfs are also special because they are close enough to
allow precise measurements of their star formation histories,
detailed spatial and kinematic structures, and to measure the
metallicities and ages of individual stars.

Over the past few years, theoretical and observational studies of
Milky Way dwarf galaxies have flourished as a result of new interest
motivated by the 'missing galaxy' and the 'cusp/core' problems with
CDM cosmologies, as well as improvements in observational and
computational resources.  The Milky Way satellites are now being
studied with unprecedented detail
(\citealt{palma03,tolstoy04,wilkinson04,babusiaux05,mayer05,munoz05},
among many others), and new satellites and remnants thereof are being
discovered around both the Milky Way and M31 (\citealt{newberg02,
yanny03,
ibata03,rochapinto04,majewski03,martin04,zucker04},
among many others).

Although systematic searches have successfully identified some of the
Milky Way dwarfs, there currently aren't well-defined, quantitative
limits on the faint end of the local galaxy luminosity function.  The
possibility also remains that existing survey data has not yet been
searched to the faintest possible depths for new dwarf galaxies.  This
lack of a well-defined sample of dwarfs currently undermines our
understanding of the ``substructure problem'': that cold dark matter
(CDM) cosmologies predict more than an order of magnitude more low
mass dark matter halos than the number of dwarfs observed around
galaxies such as the Milky Way \citep{klypin99,moore99}.  Models that
implement baryonic physical processes into CDM models of galaxy
formation have made new predictions for the observable population of
dwarfs around the Milky Way and M31
\citep{benson02LG,kravtsov04}. However any comparison between the
observed dwarf populations and these predictions is rendered
less meaningful by the uncertain completeness of the local dwarf
galaxy population \citep{willman04}.

To create a well-defined census of Milky Way dwarfs to fainter limits
than previously possible, we have been conducting a uniform, automated
search for new Milky Way companions \citep{willman02}.  Two companions
have been discovered as a result of this search thus far.

\section{An SDSS Survey For Milky Way Companions}

To identify candidates for Milky Way companions, we search for
statistical fluctuations in star counts in the Sloan Digital Sky
Survey catalog (SDSS; \citealp{york00}). SDSS data is reduced with an
automatic pipeline consisting of: astrometry \citep{pier03}; source
identification, deblending and photometry \citep{lupton01};
photometricity determination \citep{hogg01}; calibration
\citep{fukugita96,smith02}; and spectroscopic data processing
\citep{edr}. 

Our search algorithm enhances the apparent overdensity of an
extragalactic stellar system over the foreground of the Milky Way by
combining color and magnitude cuts on resolved stars with spatial
smoothing. Candidates can be identified either with or without a
color-cut designed to select stars with $g-r$, $r-i$ colors
consistent with those of metal-poor red giant branch stars.  Although
it is somewhat crude, this technique is sensitive to dwarfs many times
fainter than those known.  Our
primary survey only includes stars brighter than $r$ = 21.5 so that
star-galaxy separation is uniform enough to produce well-defined
detection limits.  However, we also extend our analysis to stars with
fainter magnitudes to maximize the possibility of finding new
companions.

The detection limits of the primary survey are a function of: i) the
density of the stellar foreground due to the Milky Way and ii) the
stellar surface density as a function of color and magnitude of a
candidate dwarf (which is a function of the dwarf's central surface
brightness, scale size, distance, star formation history, and
metallicity).  First, we determine the detection thresholds that
produce only $\sim$ 1-2 spurious detections in 1000 deg$^2$ of
randomly distributed stars.  We then use those adopted thresholds to
calculate the survey's detection limits. We simulate the stellar
surface densities of metal-poor, old population dwarf spheroidal
galaxies with a range of sizes and distances, based on a template
stellar luminosity function created with SDSS observations of Palomar
5.  The simulated galaxies are embedded in SDSS data at a range of
stellar foreground densities and the resulting detection efficiencies
determine the detection limits. See Willman et al. (2002) and Willman
et al (2005, in prep) for a detailed description of the survey
technique and detection limits.

Figure 1 shows the limiting absolute magnitude for a fiducial
direction of (l,b) = (0,50).  These limits are updated from the
preliminary results published in \citet{willman02}.  Each line in this
Figure corresponds to the absolute magnitude corresponding to a 50\%
detection efficiency for a galaxy of a different physical size.  The
magnitudes and scale sizes of the known Milky Way dwarf spheroidals
are overplotted.  This Figure shows that the survey is sensitive to
galaxies much fainter than any yet known (not including Ursa Major) to
distances beyond the Milky Way's virial radius.  We have verified that
all of the Milky Way dwarfs imaged by SDSS are detected at many
$\sigma$ over the search's detection threshold.  The (very uncertain)
magnitude and scale size of the Ursa Major dwarf galaxy (UMa), the
first new companion produced by the primary survey, is also
overplotted.

\begin{figure}[!h]
\plotone{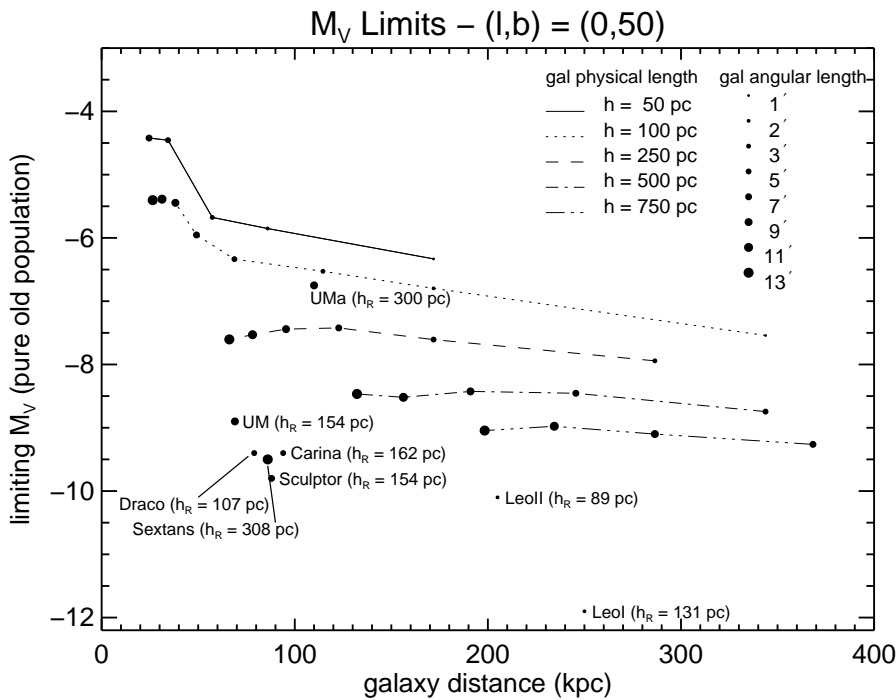}
\caption{Absolute magnitude limits of our survey for a fiducial
direction as a function of dwarf size and distance. These limits were
calculated using the stellar luminosity function of Palomar 5 and
assuming a purely old stellar population (see Willman et al 2002 and
Willman et al 2005, in prep for details).  The physical scale lengths and
absolute magnitudes of the known Milky Way dSph companions are
overplotted (from \citealt{grebel03} and Willman et al. 2005b),
although the plotted values for Ursa Major are quite uncertain.  Dwarf
galaxies several times fainter than any known (not including Ursa
Major) are detectable within 350 kpc.}
\end{figure}

\section{The Ursa Major Dwarf}
 
Our primary survey has included $\sim$ 4700 deg$^2$ of sky thus far
and has produced 17 candidates, not including all of the previously
known Milky Way companions that were detected.  The first candidate we
obtained follow-up imaging of was detected as an 8.5$\sigma$
fluctuation in the number of red stars with $19.0 < r < 20.5$ and is
located in the Ursa Major constellation at ($\alpha_{2000},
\delta_{2000}$) $=$ (158.72,51.92). The distribution of stellar
densities produced by our algorithm is not quite Gaussian, so
8.5$\sigma$ is actually only 0.3$\sigma$ above our detection
threshold.

Figure 2 shows the SDSS color-magnitude diagram of Ursa Major
alongside the SDSS color-magnitude diagram (CMD) of Sextans, one of
the two lowest surface brightness galaxies known prior to Ursa Major.
Sextans is an old and metal-poor ([Fe/H] = -2.1 $\pm$ 0.3;
\citealt{lee03}) Milky Way dSph at a distance of 86 kpc. The stellar
population of Ursa Major (UMa) is strikingly similar to that of
Sextans, including the morphologies of their horizontal and red giant
branches, suggesting they may have similar ages and metallicities.
UMa also has roughly an order of magnitude fewer stars than Sextans,
which is remarkable given that the surface brightness of Sextans is
only $\mu_V = 26.2$.

\begin{figure}[!h]
\plotone{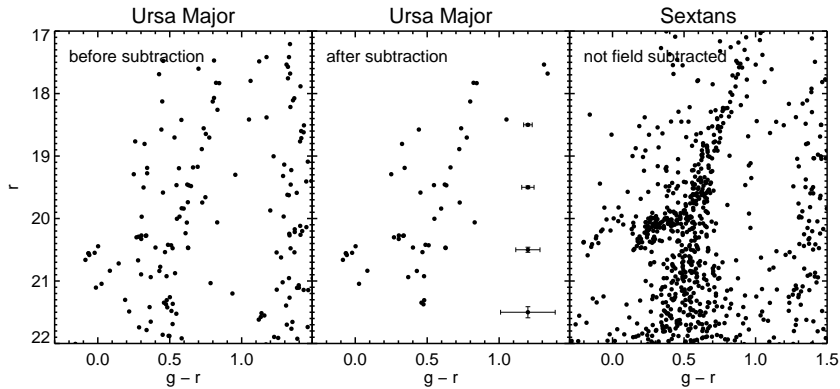}
\caption{ From \citet{willman05b}. {\it Left Panel:} Color-magnitude
   diagram including all 172 stars within the 200 arcmin$^2$ area
   included in the Ursa Major detection, without a statistical
   subtraction of foreground stars. {\it Middle Panel:} Field
   subtracted CMD of UMa. {\it Right Panel:} The CMD of all stars
   within the half-light radius of the Sextans dSph ($\mu_V \!= 26.2$,
   $d\! = 86$ kpc) without any field star subtraction. All three CMDs
   and the field subtraction were created solely with SDSS data.}
\end{figure}

Follow-up imaging obtained at the Isaac Newton Telescope in March 2005
in B and $r$ revealed that this detected overdensity truly is a dwarf
galaxy composed of an old stellar population at a distance of $\sim$
100 kpc.  We used the DAOPHOT II/ALLSTAR package \citep{stetson94} to
obtain photometry of the resolved stars. Figure 3, from \citet{willman05b}, shows that an [Fe/H]
= -1.7, 13 Gyr isochrone projected to 100 kpc provides a good match to
Ursa Major's INT color-magnitude diagram.

In \citet{willman05b}, we estimated some of Ursa Major's properties
and compared them to other known systems.  By a comparison with the
stellar luminosity functions of Sextans and Palomar 5, we estimated
the absolute magnitude of Ursa Major to be M$_V$ $\sim$ -6.75, which is
several times fainter than the faintest dwarf previously known.  Based
on the spatial distribution of red giant branch stars in the Ursa
Major dwarf, and assuming a distance of 100 kpc, we estimated
$r_{half-light,UMa} \sim$ 250 pc, which is similar to the half-light
radius of Sextans (200 pc).  This absolute magnitude and size are
currently quite uncertain, but do give a sense of UMa's properties
relative to other known systems.  Both its combination of a faint
total luminosity with a relatively large half-light radius and the
fact that UMa is more distant than all but one of the Milky Way's
150$+$ globular clusters, cause us to conclude that Ursa Major is the
tenth confirmed dwarf spheroidal companion to the Milky Way (see below
for some additional discussion).

\begin{figure}[!h]
\plotone{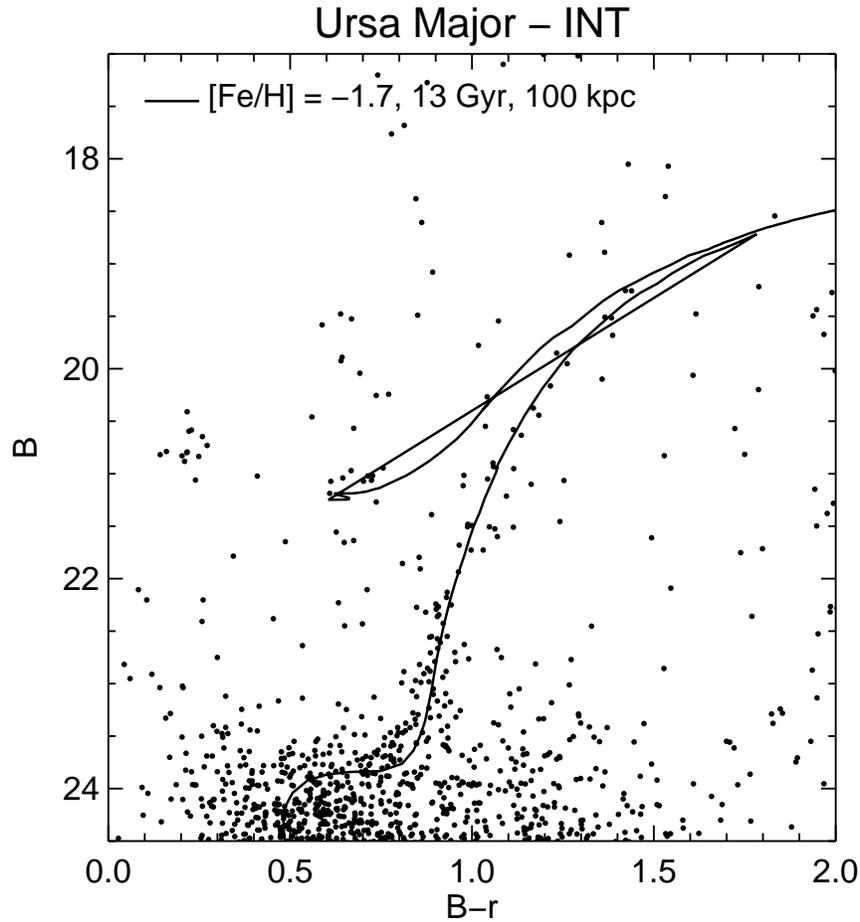}
\caption{From \citet{willman05b}.  The CMD of stars in a 23$'$
$\times$ 12$'$ field around the center of Ursa Major, as observed in a
total of 5600 seconds of exposure time in B and 4800 seconds in r.  A
theoretical isochrone of an [Fe/H] = -1.7, 13 Gyr old population
projected to 100 kpc is overplotted.  \citep{girardi04}}
\end{figure}

\section{SDSSJ1049+5103}

An extended survey analysis including stars as faint as $r$ = 23.0
produced numerous candidates, many of which appear to be cluster
galaxies misclassified as stars or globular clusters around nearby
galaxies that get classified as stars.  However one system,
SDSSJ1049+5103, stood out as a strong candidate for a new Milky Way
companion.  In \citet{willman05a}, we used SDSS data to determine that
this companion is an old, metal-poor stellar population at d $\sim$ 50
kpc, with M$_V$ $\sim$ -3, and r$_{half-light}$ $\sim$ 25 pc.

We then obtained follow-up deep, wide-field imaging of SDSSJ1049+5103
at the INT in March 2005 and again used the DAOPHOTII/ALLSTAR package
to obtain photometry of stellar sources.  The right panel of Figure 4
shows the CMD of stars from chip 4 of the WFC that lie within 3.5$'$
(2r$_{half}$) of the center of SDSSJ1049+5103, with boxes outlining
the main-sequence turnoff and the sub-giant branch. This CMD extends
almost two magnitudes fainter than the main sequence turnoff. When
extended to brighter magnitudes, this CMD shows few possible
horizontal branch or red giant branch stars (see
\citealt{willman05a}).  Despite SDSSJ1049+5103 having an ultra-low
luminosity, its stellar population clearly dominates the stars in this
CMD. However, the CMD of field stars observed on chips 1, 2, and 3 and
shown in the left panel of Figure 4 is dominated by field stars.
These adjacent chips do not display a clear signature of object stars,
but there is a hint that some SDSSJ1049+5103 stars may be found in
these chips. This hint suggests that SDSSJ1049+5103 may be more
extended than originally thought or may be getting tidally stripped.

\begin{figure}[!h]
\plottwo{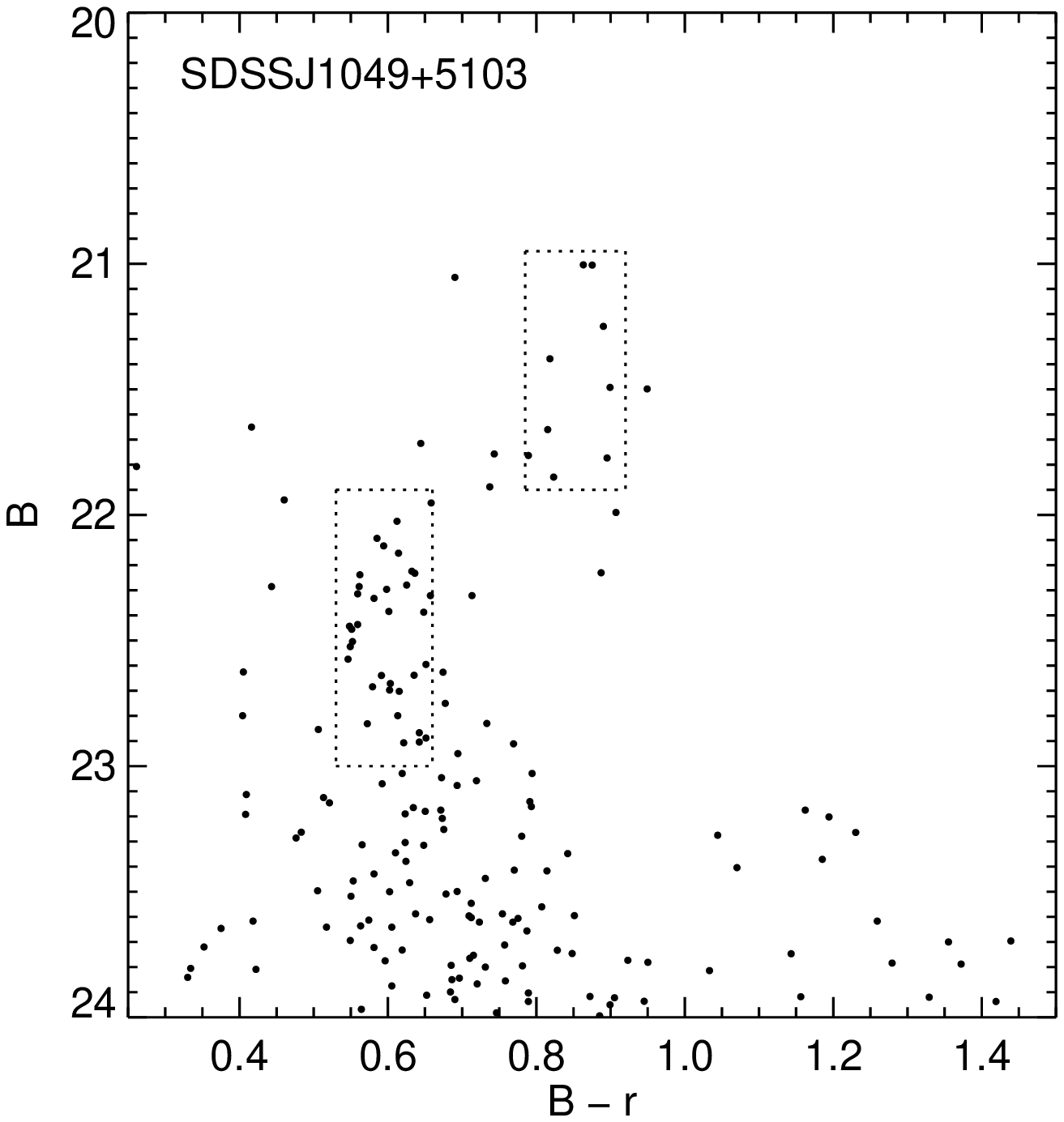}{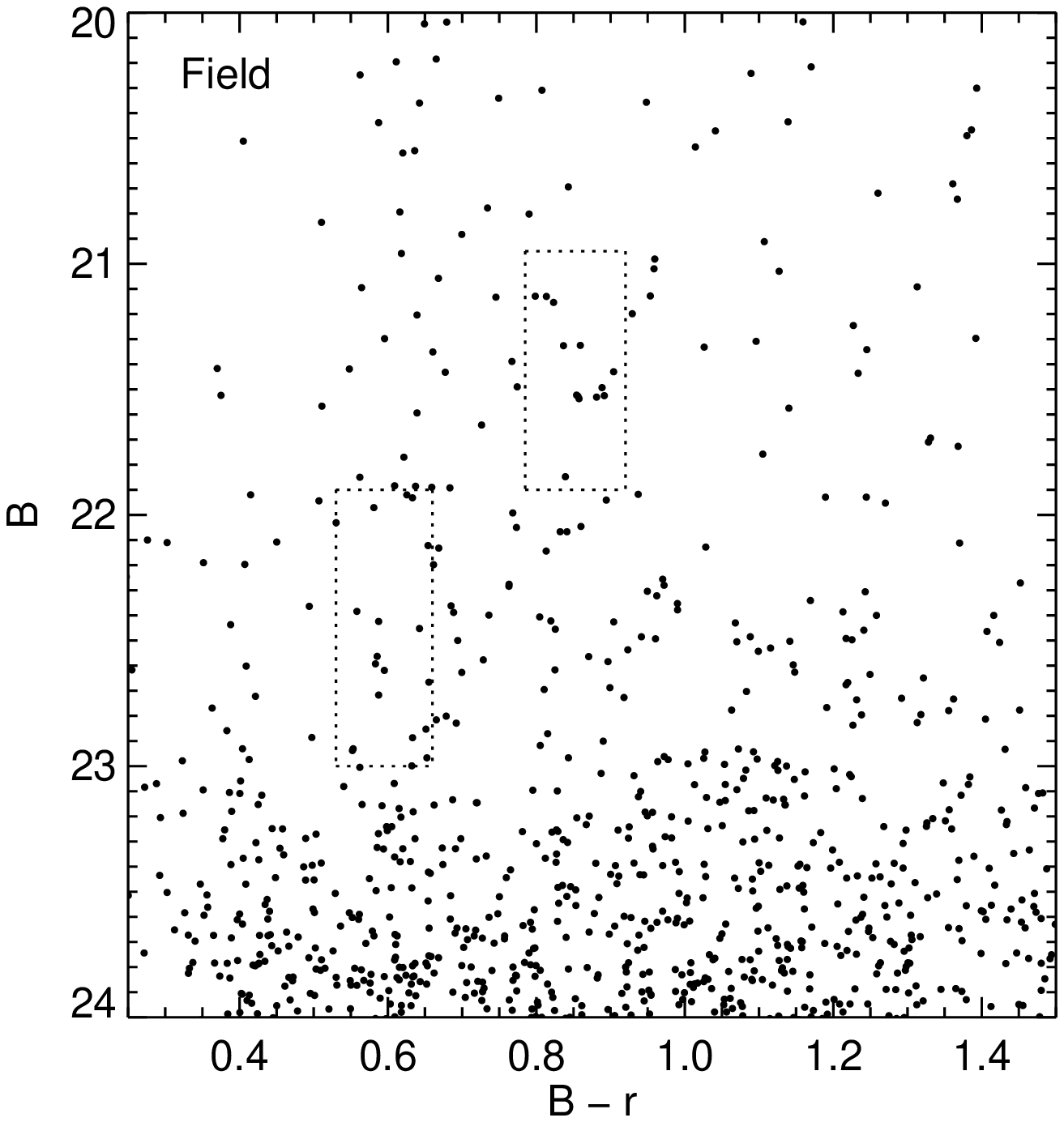}
\caption{{\it Right Panel:} CMD of stars within 3.5$'$ of the center
of SDSSJ1049+5103 (from chip 4 of the WFC on the INT). The
main-sequence turnoff and sub-giant branch are outlined.  {\it Left
Panel:} CMD of stars in surrounding fields (a total of $\sim$800$^2$;
all stars on chips 1, 2 and 3).}
\end{figure}

Figure 5 also suggests the possibility of tidal stripping.  The
spatial distribution of stars in the main-sequence turnoff and
sub-giant branch boxes is not symmetric and clearly displays a
``tail'' extending to the west of the primary object.  The entire area
of this figure is covered by chip 4 of the INT observation.  We will
present a more detailed analyses of these data in an upcoming paper.

\begin{figure}[!h]
\plotone{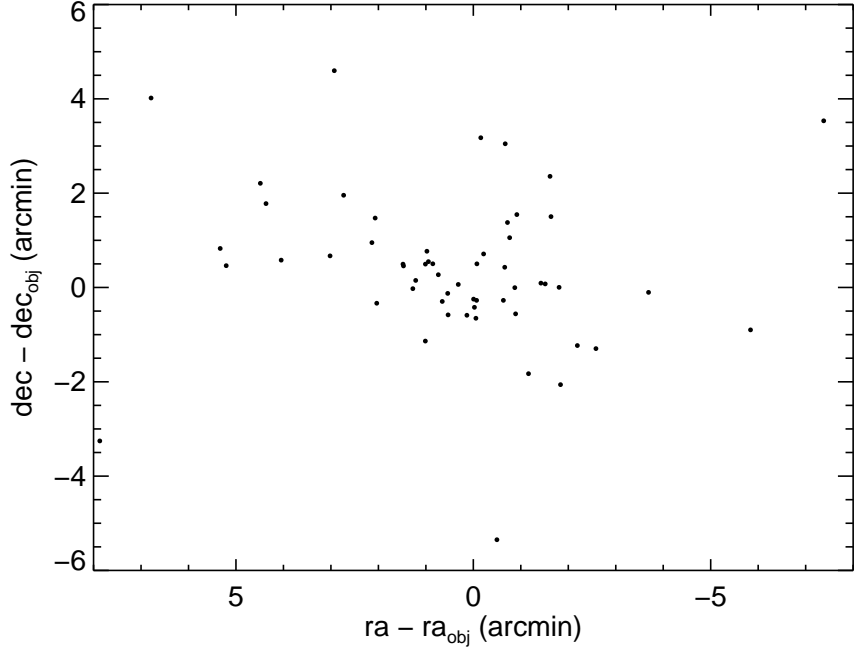}
\caption{The spatial distribution of stars centered on SDSSJ1049+5103.
This figure includes all stars that fall in the main-sequence turnoff
and sub-giant branch boxes plotted on the color-magnitude diagrams of
Figure 4.}
\end{figure}

\section{SDSSJ1049+5103 and Ursa Major: globular clusters or dwarf galaxies?}

We have shown that Ursa Major and SDSSJ1049+5103 are newly discovered
Milky Way companions.  The fact that their luminosities overlap those
of known globular clusters more than those of known dwarf galaxies
raises the question: Are they star clusters or dwarf galaxies?  The
likely fundamental difference between globular clusters and dwarf
galaxies is that a dwarf galaxy forms inside of its own dark matter
halo and a globular cluster does not.  However, the most reliable way
to observationally classify such objects has been the fact that
globular clusters are much more compact than dwarfs at a given
luminosity.

Figure 6 displays the absolute magnitude vs. half light size of: 1)
Milky Way globular clusters, 2) Milky Way dwarf spheroidals, 3) faint
red galaxies \citep{blanton04}, 3) And IX and Ursa Major dSph, and 4)
SDSSJ1049+5103. At magnitudes brighter than $M_V$ = -7, the globular
cluster (GC) and dwarf galaxy populations clearly separate in this
plot (although see \citealp{huxor05}).  However, the size-luminosity
relationships of globular clusters and of Milky Way dSphs overlap at
low luminosities. This overlap highlights the vague distinction
between these two classes of objects and shows that this simple
classification scheme is no longer sufficient, now that searches are
uncovering companions at such low luminosities. Indeed,
\citet{benson02LG} predict the existence of Milky Way dwarf satellite
galaxies as faint as the faintest GCs and with half mass radii that
roughly follow the same luminosity-size relation as the known dSphs.

\begin{figure}[!h]
\plotone{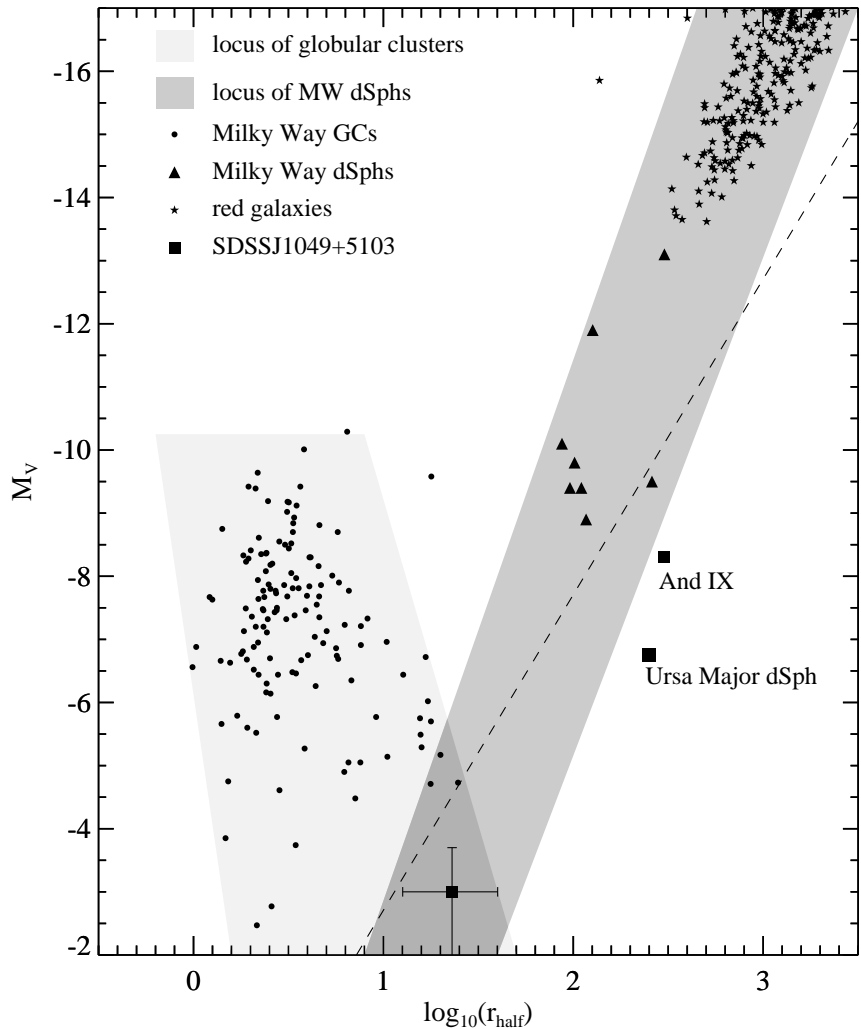}
\caption{The absolute magnitudes and half-light radii of Milky Way
globular clusters (circles), MW dwarf spheroidal galaxies (triangles),
faint red galaxies in the SDSS (stars; \citealp{blanton04}), And IX
\citep{zucker04}, Ursa Major dSph \citep{willman05b}, and
SDSSJ1049+5103 (square).  The globular cluster AM 4 is too faint
(M$_V$ = +0.2) to be included on this plot. The approximate loci of
the globular cluster and the dwarf spheroidal data are shaded.  The
Milky Way dSphs appear to follow a similar size-luminosity relation as
other faint red galaxies. A fiducial line of constant $\mu_{50}$ is
also overplotted for reference.  GC and MW dSph data are from
\citet{harrisGCcat}, \citet{mateo98}, and \citet{grebel03}.}
\end{figure}

And IX and Ursa Major both fall close to the size-magnitude
relationship followed by dwarf galaxies, but quite far from that
followed by GCs.  It is possible that they lie to the right of the
current dwarf locus simply because dwarfs in that region of
size-magnitude space are too low surface brightness to have been
detected previously. The dashed line shows a fiducial line of constant
$\mu_{50}$ for comparison.

Unlike Ursa Major, SDSSJ1049+5103 lies at the intersection of the
globular clusters and the dwarf galaxies.  Although many times fainter
than the faintest dwarf, SDSSJ1049+5103 is also $\geq$ 5 times larger
in physical size than similarly faint GCs.  It thus remains unclear
whether it is a globular cluster or a dwarf galaxy.  It may be
possible to use tidal features around SDSSJ1049+5103 to constrain its
current dark matter content \citep{moore96}, which may or may not
reflect the conditions under which it formed.

\section{Summary and Future Directions} 

The discoveries of at least one new Milky Way satellite and at least
one other ambiguous companion in $<$ 1/8 of the sky suggests that many
more ultra-faint Milky Way satellites may yet to be discovered. These
discoveries also raise a number of interesting questions whose answers
could impact our global understanding of galaxy formation: Do these
systems have few stars as a result of nature or nurture?  What is the
lower limit of galaxy formation?  What is the relationship between
globular clusters and dwarf galaxies?

We are in the process of obtaining and analyzing follow-up imaging for
the remaining 16 candidates produced by our survey.  We are also
currently performing a more detailed analysis of deep
imaging of both Ursa Major and SDSSJ1049+5103, as well as of HIRES
spectra of some Ursa Major stars.

\acknowledgements I thank David Valls-Gabaud, Miguel Chavez, and
everyone else on the LOC and SOC for bringing this conference
together.  I also thank and acknowledge my collaborators Julianne
Dalcanton, David Martinez-Delgado, Michael Blanton, Andrew West, and
David Hogg for their contributions to the research presented in this
proceeding.


\end{document}